\documentclass[twocolumn]{revtex4-1}
\usepackage[dvipdfmx]{graphicx}
\usepackage{amsmath,amssymb,bm,amsthm}
\usepackage{color}
\usepackage{ulem}
\usepackage{here}

\newcommand{\nc}{\newcommand}
\nc{\rnc}{\renewcommand}
\nc{\nn}{\nonumber}

\begin{document}
\title{Entanglement prethermalization in the Tomonaga-Luttinger model}

\author{Eriko Kaminishi$^1$, Takashi Mori$^1$, Tatsuhiko N. Ikeda$^{2}$ and Masahito Ueda$^{1,3}$}
\affiliation{
$^1$Department of Physics, University of Tokyo, \\ 7-3-1 Hongo, Bunkyo-ku, Tokyo 113-0033, Japan\\
$^2$Institute for Solid State Physics, University of Tokyo, Kashiwa, Chiba 277-8581, Japan\\
$^3$RIKEN Center for Emergent Matter Science (CEMS), Wako, Saitama 351-0198, Japan
} 
\date{\today}

\begin{abstract}
{Prethermalization refers to the relaxation to a quasi-stationary state before reaching thermal equilibrium.
Recently, it is found that not only local conserved quantities but also entanglement  plays a key role in a special type of prethermalization, called entanglement prethermalization.
Here, we show that in the Tomonaga-Luttinger model the entanglement prethermalization can also be explained by the conventional prethermalization of two independent subsystems without entanglement.
Moreover, it is argued that prethermalization in the Tomonaga-Luttinger model is essentially different from entanglement prethermalization in the Lieb-Liniger model 
because of the different types of energy degeneracies. 
}
\end{abstract}
\maketitle

\section{Introduction}
Relaxation and thermalization in an isolated quantum system have attracted growing interest from the viewpoint of fundamental principles of quantum and statistical physics \cite{Neumann1929, Srednicki1994,  Tasaki1998, Popescu2006, Rigol2008, Linden2009, Sato2012} and experimental investigations using cold atoms \cite{Kinoshita2006, Hofferberth2007, Trotzky2012, Kaufman2016}.
An isolated system often relaxes to a quasi-stationary state, a phenomenon known as prethermalization \cite{Berges2004, Kollar2011, Gring2012, Langen2013ST, Langen2013, Kuhnert2013, Worm2013, Langen2015}.
It is known that prethermalization usually occurs due to the existence of local quantities which are almost conserved over a certain long timescale.
In some idealized case of, e.g., an integrable system, the system never thermalizes, and a non-thermal steady state defined in such an idealized limit corresponds to a quasi-stationary prethermalized state in more realistic situations.
Because of small deviations from the idealized limit, the system will eventually reach thermal equilibrium.
In this work, we do not consider such small deviations from the idealized limit, and hence we here treat only the first stage of the relaxation to a prethermalized state.

It has recently been found that the initial entanglement between two subsystems can affect prethermalization, which we call entanglement prethermalization (EP) \cite{Kaminishi2015, Ikeda2017}.
EP has been demonstrated in the Lieb-Liniger (LL) model \cite{Lieb1963}, which describes a one-dimensional Bose gas with a contact interaction.
We prepare the ground state of the LL gas and split it into two independent subsystems, let them evolve in time, and finally measure the interference pattern of the two overlapping Bose gases.
It is found that the cross-correlation function of the two subsystems depends on the initial entanglement between them even in the long-time limit \cite{Kaminishi2015}.

On the other hand, prethermalization in the interference pattern between the split one-dimensional Bose gases has been experimentally observed \cite{Gring2012}, and theoretically analyzed by using the Tomonaga-Luttinger (TL) model \cite{Gring2012, Kitagawa2010, Kitagawa2011}.
Since the TL model is the low-energy effective theory of the LL model, one may ask whether EP found in the LL model is identical to the prethermalization theoretically analyzed in the TL model.

In this paper, we show that the prethermalization observed in the split TL gases can also be interpreted as EP between the two subsystems, but that it should be distinguished from that in the LL model.
The nature of energy degeneracies plays a key role here as explained in Sec.~\ref{sec:EP}.
The TL model can be mapped to a system of non-interacting bosons and there are many degeneracies in the Hamiltonian.
On the other hands, in the LL model, many energy degeneracies present in the TL model are lifted due to the nonlinearity of the interaction.
The remaining energy degeneracies due to the translation symmetry and the inversion symmetry cause the EP in the LL model.
This difference leads to the distinction concerning the EP between the TL model and the LL model.

The rest of this paper is organized as follows.
In Sec. II, we explain general mechanism of EP in two noninteracting subsystems under a unitary time evolution.
In Sec. III, we briefly explain the known result on the EP in the LL model. 
In Sec. IV, we study the EP in the TL model.
In Sec. V, we consider the case in which there are interactions between two subsystem.
In Sec. VI, we summarize the mail results of this paper.

\section{Entanglement prethermalization}
\label{sec:EP}
We first briefly explain general mechanism of entanglement prethermalization (EP) in two noninteracting subsystems under the unitary time evolution.
Usually, the initial energy is relevant for the long-time behavior of physical quantities, but the initial entanglement is not.
However, if there are some energy degeneracies, the entanglement survives and can make significant contributions to the long-time behavior of physical quantities.

Here, we remark that we discuss the prethermalized state reached after the first relaxation, and do not consider the second relaxation due to small perturbations.
For this purpose, we only consider the infinite-time average of physical quantities without perturbations.

Two conditions are necessary for EP.
One is to prepare an initial state in which two non-interacting subsystems are entangled.
It is realized by a coherent splitting of the system as in the experiment \cite{Gring2012} .
Then, to protect the entanglement for a long time, energy degeneracies are necessary.
If these two conditions are satisfied, the long-time average of the density matrix is given by a mixture of entangled states.

To be concrete, let us consider the system consisting of the two subsystems 1 and 2.
The Hilbert space is given by $\mathcal{H}=\mathcal{H}_1\otimes\mathcal{H}_2$, where $\mathcal{H}_{\alpha}$ is the Hilbert space of the $\alpha$th subsystem ($\alpha=1,2$).
The Hamiltonian is given by $\hat{H}=\hat{H}_1\otimes\hat{1}+\hat{1}\otimes\hat{H}_2$, where $\hat{H}_{\alpha}$ represents the Hamiltonian of the $\alpha$th subsystem, and $\hat{1}$ is the identity operator.
An energy eigenstate is denoted by $|n,i;m,j\rangle=|n,i\rangle\otimes|m,j\rangle$, where $H_1|n,i\rangle=E_n^{(1)}|n,i\rangle$ and $H_2|m,j\rangle=E_m^{(2)}|m,j\rangle$.
The degree of degeneracies of the eigenvalue $E_n^{(\alpha)}$ of $H_{\alpha}$ is denoted by $d_n^{(\alpha)}$, and thus $i=1,2,\dots,d_n^{(1)}$ and $j=1,2,\dots,d_m^{(2)}$.
When the two subsystems are identical, $E_n^{(1)}=E_n^{(2)}$ and $d_n^{(1)}=d_n^{(2)}$, but we consider a more general case.
We set $\hbar=1$ throughout this paper.

Let us define the projection operator $\mathcal{P}_E$ onto the subspace with the total energy $E$,
\begin{equation}
\mathcal{P}_E\equiv\sum_{\substack{(n,i),(m,j) \\ E_n+E_m=E}}|n,i;m,j\rangle\langle n,i;m,j|.
\end{equation}
The infinite-time average of the density matrix $\rho(t)=|\Psi(t)\rangle\langle\Psi(t)|$, where $|\Psi(t)\rangle$ is the state of the total system evolving in time as $id|\Psi(t)\rangle/dt=H|\Psi(t)\rangle$, is given by
\begin{equation}
\bar{\rho}\equiv\lim_{T\rightarrow\infty}\frac{1}{T}\int_0^Tdt\rho(t)=\sum_Ep_E|\Psi_E\rangle\langle\Psi_E|,
\end{equation}
where $p_E=\langle\Psi(0)|\mathcal{P}_E|\Psi(0)\rangle$ and
\begin{align}
|\Psi_E\rangle&=\frac{1}{\sqrt{p_E}}\mathcal{P}_E|\Psi(0)\rangle
\nonumber \\
&=\frac{1}{\sqrt{p_E}}\sum_{\substack{(n,i),(m,j) \\ E_n+E_m=E}}C_{(n,i),(m,j)}|n,i;m,j\rangle,
\label{eq:Psi_E1}
\end{align}
with $C_{(n,i),(m,j)}=\langle n,i;m,j|\Psi(0)\rangle$.
We note that the decoherence occurs only between states with different eigenenergies and that the initial coherence within the degenerate subspace is maintained upon the infinite-time average.

First, we consider the case in which the condition $E_n+E_m=E$ uniquely determines $n$ and $m$.
In this case,
\begin{equation}
|\Psi_E\rangle=\frac{1}{\sqrt{p_E}}\sum_{i=1}^{d_n^{(1)}}\sum_{j=1}^{d_m^{(2)}}C_{(n,i),(m,j)}|n,i\rangle\otimes|m,j\rangle.
\label{eq:Psi_E2}
\end{equation}
If there is no initial entanglement, $C_{(n,i),(m,j)}=C_{(n,i)}^{(1)}C_{(m,j)}^{(2)}$ and
\begin{equation}
|\Psi_E\rangle=\frac{1}{\sqrt{p_E}}\left(\sum_{i=1}^{d_n^{(1)}}C_{(n,i)}^{(1)}|n,i\rangle\right)
\otimes\left(\sum_{j=1}^{d_m^{(2)}}C_{(m,j)}^{(2)}|m,j\rangle\right),
\end{equation}
and there is no entanglement in $|\Psi_E\rangle$ and in $\bar{\rho}$.
On the other hand, if the entanglement is present in the initial state, $|\Psi_E\rangle$ in (\ref{eq:Psi_E2}) is also an entangled state, and thus the initial entanglement persists during the time evolution.

When two subsystems are identical and mappable to non-interacting bosons (as in the TL model) or fermions, there will be many $n$ and $m$ satisfying $E_n+E_m=E$ for a fixed value of $E$.
Such additional energy degeneracies also contribute to the entanglement in the state $|\Psi_E\rangle$ in (\ref{eq:Psi_E1}).

In this way, the presence of the initial entanglement and the energy degeneracies provide an essential mechanism of the EP.

\section{Entanglement Prethermalization in the Lieb-Liniger model}
\label{sec:LL}
We apply the mechanism of EP to the LL model.
The initial state is prepared in the ground state of the LL Hamiltonian.
Therefore the system has translation symmetry which leads to the momentum conservation, and inversion symmetry which leads to energy degeneracy.
As a consequence, the infinite-time average of the density matrix is block-diagonalized in terms of these entangled states.

The LL Hamiltonian in the periodic boundary condition is
\begin{align}
\hat{H}^{(LL)}&=\int_{-L/2}^{L/2}dx\left({\partial_x}\hat{\psi}^{\dagger}(x){\partial_x}\hat\psi(x)\right. \nonumber\\
&\quad\left.+g \hat\psi^{\dagger}(x)\hat\psi^{\dagger}(x)\hat\psi(x)\hat\psi(x)\right),
\label{eq:LL}
\end{align}
where $\hat{\psi}(x)$ is the bosonic field operator and we employ a system of units with $2m=1$, where $m$ is the mass of the boson, and $N/L=1$.
We consider the repulsive interactions, $g>0$.
This model is integrable and exact many-body energy eigenstates can be obtained by the Bethe-Ansatz method, where the $N$-body eigenstate $|{\bm{k}}^N\rangle$ is characterized by a set of quasi-momenta ${\bm{k}}^N$ which give the eigenenergy $E({\bm{k}})=\sum_{j=1}^Nk_j^2$ and eigenmomentum $P({\bm{k}}^N)=\sum_{j=1}^Nk_j$.

To prepare an initial state with two entangled subsystems, we consider a quantum quench, which mimics a coherent split of a 1D Bose gas into the ``up'' component and the ``down'' component.
We assume that the excitation energy along the radial direction is very large.
In this process, each boson operator $\hat{\psi}(x)$ becomes a symmetric superposition, $\hat{\psi}(x)=\hat{\psi}_c(x)\equiv [\hat{\psi}_{\uparrow}(x)+\hat{\psi}_{\downarrow}(x)]/\sqrt{2}$, and there is no ``antisymmetric boson'' $\hat{\psi}_s^{\dagger}(x)\hat{\psi}_s(x)=0$, where $\hat{\psi}_s(x)\equiv [\hat{\psi}_{\uparrow}(x)-\hat{\psi}_{\downarrow}(x)]/\sqrt{2}$~\cite{Kaminishi2015}.

Therefore, the initial state is given by the ground state of
\begin{align}
\hat{H}_0^{(LL)} =\int_{-L/2}^{L/2}dx\left({\partial_x}\hat{\psi}_c^{\dagger}(x){\partial_x}\hat\psi_c(x)\right. \nonumber \\
\left.+g \hat\psi_c^{\dagger}(x)\hat\psi_c^{\dagger}(x)\hat\psi_c(x)\hat\psi_c(x)\right),
\label{LL_H0}
\end{align}
subject to the condition
\begin{equation}
\hat{\psi}_s(x)|\Psi_s(0)\rangle=0.
\label{initial}
\end{equation}

After the split, the up and down gases do not interact with each other, so the Hamiltonian after the quench is given by 
\begin{align}
\hat{H}_1^{(LL)}=\hat{H}_{\uparrow}^{(LL)}+\hat{H}_{\downarrow}^{(LL)}
\label{quench_LL}
\end{align}
where

\begin{align}
\hat{H}_{{\uparrow}/{\downarrow}}^{(LL)}&=\int_{-L/2}^{L/2}dx\left({\partial_x}\hat{\psi}_{{\uparrow}/{\downarrow}}^{\dagger}(x){\partial_x}\hat\psi_{{\uparrow}/{\downarrow}}(x)\right.\nonumber \\
&\left. \quad +g' \hat\psi_{{\uparrow}/{\downarrow}}^{\dagger}(x)\hat\psi_{{\uparrow}/{\downarrow}}^{\dagger}(x)\hat\psi_{{\uparrow}/{\downarrow}}(x)\hat\psi_{{\uparrow}/{\downarrow}}(x) \right).
\label{Hamiltonian_updown}
\end{align}
The interaction after the quench $g'$ is arbitrary, but we chose $g'=g$ in this paper.

The initial sate $|\Psi(0)\rangle$ is the ground state of $\hat{H}_0$, and the state at time $t$ is given by $|\Psi(t)\rangle={\rm{e}}^{-i\hat{H}_1t}|\Psi(0)\rangle$.
Let  $\bm{k}_{\rm {\uparrow}}^M$ and $\bm{k}_{\rm {\downarrow}}^{N-M}$ be the quasi-momenta of the up and down components, respectively.
By expanding $|\Psi(0)\rangle$ in the basis of eigenstates of $\hat{H}_1$, we obtain 
\begin{equation}
\begin{split}
|\Psi(t)\rangle=\sum_{M=0}^{N}\sum_{\bm{k}_{\rm {\uparrow}}^M,\bm{k}_{\rm {\downarrow}}^{N-M}}&C(\bm{k}_{\rm {\uparrow}}^M,\bm{k}_{\rm {\downarrow}}^{N-M})e^{-iE (\bm{k}_{\rm {\uparrow}}^M,\bm{k}_{\rm {\downarrow}}^{N-M})t}\\
&\times |\bm{k}_{\rm {\uparrow}}^M\rangle|\bm{k}_{\rm {\downarrow}}^{N-M}\rangle,
\end{split}
\label{eq:state}
\end{equation}
where
\begin{equation}
\hat{H}_1|{\bm{k}_{\rm {\uparrow}}}^M\rangle|{\bm{k}}_{\rm {\downarrow}}^{N-M}\rangle=E ({\bm{k}}_{\rm {\uparrow}}^M,{\bm{k}}_{\rm {\downarrow}}^{N-M})|{\bm{k}_{\rm {\uparrow}}}^M\rangle|{\bm{k}}_{\rm {\downarrow}}^{N-M}\rangle
\end{equation}
with $E ({\bm{k}}_{\rm {\uparrow}}^M,{\bm{k}}_{\rm {\downarrow}}^{N-M}) \equiv E ({\bm{k}}_{\rm {\uparrow}}^M)+E ({\bm{k}}_{\rm {\downarrow}}^{N-M})$.
The time evolution is straightforwardly calculated once we determine the expansion coefficients $\{C(\bm{k}_{\rm {\uparrow}}^M,\bm{k}_{\rm {\downarrow}}^{N-M})\}$ since $E (\bm{k}_{\rm {\uparrow}}^M, \bm{k}_{\rm {\downarrow}}^{N-M})$ can be calculated exactly by the Bethe ansatz method.

In Ref.~\cite{Kaminishi2015}, the prethermalization is discussed by calculating the auto-correlation of the up component in the Bose gas $C_{\uparrow}(x,t)=\langle \Psi(t)|\hat{\psi}_{\uparrow}^{\dagger}(x)\hat{\psi}_{\uparrow}(0)|\Psi(t)\rangle$ and the cross-correlation between the up and down components $C_{{\uparrow} {\downarrow}}(x,t)=\langle \Psi(t)|\hat{\psi}_{\uparrow}^{\dagger}(x)\hat{\psi}_{\uparrow}(0)\hat{\psi}_ {\downarrow}^{\dagger}(0)\hat{\psi}_ {\downarrow}(x)|\Psi(t)\rangle$.
We compare the infinity-time average and the thermal average at an effective temperature.
It is numerically shown that the infinite-time average of the auto-correlation agrees with its thermal average at the effective temperature, while, as for the cross-correlation, the infinite-time average deviates from the thermal average.
Moreover, the cross-correlation function cannot be described by a Gibbs state at any temperature.

The physics behind the EP is the energy degeneracy due to symmetries. 
The LL model has translation symmetry and inversion symmetry, which lead to energy degeneracy. As a result, the infinite-time average of the density matrix is block-diagonalized in terms of these entangled states, as discussed in Sec.II.

\section{Pre-thermalization in the TL model}
As we see below, the LL model reduces to the TL Hamiltonian in the low-energy approximation and then the quench problem considered in Sec.~\ref{sec:LL} by using the LL model reduces to a quench problem of the TL Hamiltonian.
In this section, we study the low-energy approximation of the quench problem considered in Sec.~\ref{sec:LL}.

It should be emphasized that studying the long-time behavior after the quench in the TL Hamiltonian does simply \textit{not} lead to an approximation of the result in Sec.~\ref{sec:LL} obtained by considering the infinite-time average under the LL dynamics.  
An important point is that, even in the low-energy regime, the quantum dynamics under the TL Hamiltonian well approximates the original dynamics under the LL Hamiltonian only in a finite timescale.
Thus, the long-time behavior of the LL Hamiltonian can be different from that of the TL Hamiltonian, and in that case, the long-time behavior of the TL model corresponds to the behavior of the LL model in a long but intermediate timescale. 
We will be able to gain a new insight into the behavior of the LL model in an intermediate timescale after the quench by investigating the long-time behavior of the TL model.
This is why we consider the low-energy approximation of the quench dynamics in Sec.~\ref{sec:LL}.

First, we derive the TL Hamiltonian as a low-energy effective theory of the LL Hamiltonian.
We write the Bose field operator as a product of the density part and the phase part $\hat{\psi}^{\dagger}_{\alpha}(x)=\sqrt{\rho_{\alpha}+\hat{n}_{\alpha}(x)}e^{-i\hat{\theta}_{\alpha}(x)}$.
Here, $\alpha=\uparrow$ or $\downarrow$ and $\rho_{\alpha}=N_{\alpha}/L=1/2$ because we assume $N/L=(N_{\uparrow}+N_{\downarrow})/L=1$ with $N_{\uparrow}=N_{\downarrow}$.
The commutation relations are given by $[\hat{\theta}_{\alpha}(x), \hat{\theta}_{\alpha'}(x')]=0$, $[\hat{n}_{\alpha}(x), \hat{n}_{\alpha'}(x')]=0$, and $[\hat{n}_{\alpha}(x), \hat{\theta}_{\alpha'}(x')]=i\delta(x-x')\delta_{\alpha\alpha'}$ \cite{Giamarchi2004}.
Then, we consider that these fluctuations are very small which corresponds to the low-energy approximation.
After that we obtain the TL model as a low-energy effective theory of the LL model.
The TL Hamiltonian corresponding to (\ref{LL_H0}) is given by
\begin{align}
H_{0}^{(TL)}=\int_{-L/2}^{L/2}\left[\frac{(\partial_x \hat{\theta}_c(x))^2}{4}+4g\hat{n}_c^2(x)\right]dx,
\label{TL_H0}
\end{align}
and the condition (\ref{initial}) reduces, in the linear order of $\hat{n}$ and $\hat{\theta}$, to 
\begin{align}
\left(\hat{n}_s(x)+i\frac{\hat{\theta}_s(x)}{2}\right)|\Psi_s(0)\rangle=0,
\label{s_initial}
\end{align}
where the ``charge'' and ``spin'' components are defined as 
\begin{align}
&\hat{n}_c(x)=\frac{\hat{n}_{\uparrow}(x)+\hat{n}_{\downarrow}(x)}{2}, \quad
\hat{n}_s(x)=\frac{\hat{n}_{\uparrow}(x)-\hat{n}_{\downarrow}(x)}{2}, \\
&\hat{\theta}_c(x)=\hat{\theta}_{\uparrow}(x)+\hat{\theta}_{\downarrow}(x), \quad
\hat{\theta}_s(x)=\hat{\theta}_{\uparrow}(x)-\hat{\theta}_{\downarrow}(x).
\end{align}
Thus, the initial state $|\Psi(0)\rangle$ is the ground state of (\ref{TL_H0}) under the condition (\ref{s_initial}).
The TL Hamiltonian after the quench, which corresponds to (\ref{quench_LL}) and (\ref{Hamiltonian_updown}), is given by 
\begin{align}
\hat{H}_1^{(TL)}=\hat{H}_{\uparrow}^{(TL)}+\hat{H}_{\downarrow}^{(TL)}
\label{TL_1updown}
\end{align}
with 
\begin{align}
\hat{H}_{\uparrow/\downarrow}^{(TL)}=\int_{-L/2}^{L/2}\left[\frac{(\partial_x\hat{\theta}_{\uparrow/\downarrow}(x))^2}{2}+g\hat{n}^2_{\uparrow/\downarrow}(x)\right]dx.
\label{uparrow/downarrow}
\end{align}

The TL model is considered as a collection of harmonic oscillators, and Eq.~(\ref{uparrow/downarrow}) is diagonalized as 
\begin{align}
\hat{H}_{\uparrow/\downarrow}^{(TL)}=\sum_{k} \omega_k b_{k_{\uparrow/\downarrow}}^{\dagger}b_{k_{\uparrow/\downarrow}},
\label{H_b}
\end{align}
where
\begin{align}
\omega_k=|k|\sqrt{2g}, \quad k=\frac{2\pi n}{L} \quad (n=0,\pm1,\pm2,\dots),
\end{align}
\begin{align}
b_{k{\uparrow/\downarrow}}\equiv\left[\frac{1}{|k|}\sqrt{\frac{g}{2}}\right]^{\frac{1}{2}}\tilde{n}_{\uparrow/\downarrow}(k)+i\left[\frac{|k|}{2}\sqrt{\frac{1}{2g}}\right]^{\frac{1}{2}}\tilde{\theta}_{\uparrow/\downarrow}(k),
\end{align}
and
\begin{align}
\left\{
\begin{aligned}
&\tilde{n}_{\uparrow/\downarrow}(k)=\frac{1}{\sqrt{L}}\int_{-L/2}^{L/2}dxn_{\uparrow/\downarrow}(x)e^{-ikx}, \\
&\tilde{\theta}_{\uparrow/\downarrow}(k)=\frac{1}{\sqrt{L}}\int_{-L/2}^{L/2}dx\theta_{\uparrow/\downarrow}(x)e^{-ikx}.
\end{aligned}
\right.
\end{align}
The state at $t$ is given by $|\Psi(t)\rangle=e^{-i\hat{H}_1^{(TL)}t}|\Psi(0)\rangle$.

In deriving Eq.~(\ref{TL_H0}), we have assumed that $\partial_x\hat{n}(x)$ is negligible compared with $\sqrt{g}\hat{n}(x)$.
This approximation is justified for $|k|<k_c:=\sqrt{g}\sim 2\pi/\xi$, where $k_c$ is the ultraviolet cutoff and $\xi$ is the healing length~\cite{Kitagawa2011}.
Thus we should always consider the Fourier modes within the range $|k|<k_c$.

Since the time evolution is determined by the Hamiltonian (\ref{H_b}), we shall express the initial state in terms of $b^{\dagger}_{k\uparrow}$ and $b^{\dagger}_{k\downarrow}$.
From the above condition, the initial state is given by the product of the charge part and the spin part:
\begin{align}
|\Psi(0)\rangle=|\Psi_{\rm c}(0)\rangle \otimes |\Psi_{\rm s}(0)\rangle.
\label{eq:initial}
\end{align}
The charge part of the initial state $|\Psi_c(0)\rangle$ is the ground state of equation (\ref{TL_H0}), which is nothing but the two-mode squeezed vacuum in terms of $\{b_k^c,b_k^{c\dagger}\}$, where 
\begin{align}
b_k^c=\frac{b_{k\uparrow}+b_{k\downarrow}}{\sqrt 2}
\label{bkc}
\end{align}
\begin{align}
|\Psi_c(0)\rangle=\prod_{k:0<k< k_c}\frac{1}{\cosh r_c}e^{-\tanh(r_c)b_k^{c\dagger}b_{-k}^{c\dagger}}|0\rangle
\end{align}
with
\begin{align}
e^{-r_c}=\left(\frac{1}{2}\right)^{1/4},
\end{align}
and $|0\rangle$ is the vacuum, $b_{k\uparrow/\downarrow}|0\rangle=0$.
For the spin part, (\ref{s_initial}) implies
\begin{align}
|\Psi_s(0)\rangle=\prod_{k:0<k<k_c}\frac{1}{\cosh[r_s(k)]}e^{-\tanh[r_s(k)]b_k^{s\dagger}b_{-k}^{s\dagger}}|0\rangle
\label{initial_psi_S}
\end{align}
and
\begin{align}
e^{-r_s(k)}=\left(\frac{2g}{k^2}\right)^{1/4},
\end{align}
where
\begin{align}
b_k^s=\frac{b_{k\uparrow}-b_{k\downarrow}}{\sqrt 2}.
\label{bks}
\end{align}
The state of the spin component (\ref{initial_psi_S}) is the two-mode squeezed vacuum.
We note that the initial state is entangled in the up-down representation:
\begin{equation}
|\Psi(0)\rangle\neq|\Psi_{\uparrow}(0)\rangle\otimes|\Psi_{\downarrow}(0)\rangle.
\end{equation}

\subsection{EP in the TL model}
\label{sec:up_down}
First, we study the time evolution after the quench in the spin-up and spin-down representation.
By substituting (\ref{bkc}) and (\ref{bks}) to (\ref{eq:initial}), we obtain the up-down representation of the initial state which is explicitly given by
\begin{align}
|\Psi(0)\rangle=\sum_{\{ n_k\} }C_{\{n_k\}}|\Phi_{\{n_k\}}\rangle,
\end{align}
where
\begin{widetext}
\begin{align}
C_{\{n_k\}}|\Phi_{\{n_k\}}\rangle=\prod_{k:0<k<k_c}\frac{1}{\cosh r_c\cosh r_s(k)}\frac{(-1)^{n_k}}{n_k!}\left[R_+(b_{k\uparrow}^{\dagger}b_{-k\uparrow}^{\dagger}+b_{k\downarrow}^{\dagger}b_{-k\downarrow}^{\dagger})+R_-(b_{k\uparrow}^{\dagger}b_{-k\downarrow}^{\dagger}+b_{k\downarrow}^{\dagger}b_{-k\uparrow}^{\dagger})\right]^{n_k}|0\rangle
\end{align}
\end{widetext}
Here, $R_+:=(\tanh r_c+\tanh r_s(k))/2$, $R_-:=(\tanh r_c-\tanh r_s(k))/2$, and the state $|\Phi_{\{n_k\}}\rangle$ is characterized by the set of nonnegative integers $n_k$ for each mode $k$ and the normalization condition $\langle\Phi_{\{n_k\}}|\Phi_{\{n_k\}}\rangle=1$.

Since a pair of bosons with momenta $+k$ and $-k$ has the energy $2\omega_k$, $|\Phi_{\{n_k\}}\rangle$ is a superposition of degenerate energy eigenstates with energy $\sum_k 2n_k\omega_k$.
The infinite-time average of the density matrix is obtained as~\footnote
{Here, we have neglected the presence of further degeneracies, i.e., the presence of $\{n_k\}\neq\{n_k'\}$ with $\sum_k2n_k\omega_k=\sum_k2n_k'\omega_k$.
This approximation is justified when we consider the auto- and cross-correlation functions because matrix elements between those states are not relevant. 
}
\begin{align}
\bar{\rho}=\sum_{\{n_k\}}|C_{\{n_k\}}|^2|\Phi_{\{n_k\}}\rangle\langle\Phi_{\{n_k\}}|.
\label{eq:rho_bar}
\end{align}
It is noted that each $|\Phi_{\{n_k\}}\rangle$ is an entangled state of spin-up and spin-down.
There are many energy degeneracies, and thus, according to the argument in Sec. II, the influence of the initial entanglement lasts forever, and the system will reach a stationary sate which is different from thermal equilibrium.
This is nothing but EP.

If the initial state of the two subsystems (spin-up and spin-down) are prepared independently and there is no entanglement between them, each subsystem will evolve to the generalized Gibbs ensemble (GGE) \cite{Rigol2007, Rigol2006},
\begin{equation}
\rho_{\rm \uparrow \downarrow}^{\rm GGE}=\frac{\exp[-\sum_k(\lambda_{k \uparrow} b_{k \uparrow}^{\dagger}b_{k \uparrow} + \lambda_{k \downarrow} b_{k \downarrow}^{\dagger}b_{k \downarrow})]}{\mathrm{Tr}\,\exp[-\sum_k(\lambda_{k \uparrow} b_{k \uparrow}^{\dagger}b_{k \uparrow} + \lambda_{k \downarrow} b_{k \downarrow}^{\dagger}b_{k \downarrow})]},
\label{eq:rho_gge_updown}
\end{equation}
because it is known that integrable systems relax to the GGE~\cite{Cazalilla2009, Cazalilla2012, Cazalilla2008, Calabrese2011, Essler2012, Fagotti2013, Bucciantini2014, Pozsgay2013, Fagotti2014, Mierzejewski2014, Caux2012, Mossel2012, Collura2013, Pozsgay2014, Sotiriadis2014, Goldstein2015}.
Here, the parameters $\{\lambda_{k\uparrow\downarrow}\}$ are determined by the initial values of the conserved quantities
\begin{equation}
\langle\Psi(0)|b_{k\uparrow\downarrow}^{\dagger}b_{k\uparrow\downarrow}|\Psi(0)\rangle=\mathrm{Tr}\,\rho_{\rm \uparrow \downarrow}^{\rm GGE}b_{k\uparrow\downarrow}^{\dagger}b_{k\uparrow\downarrow}=\frac{1}{e^{\lambda_{k\uparrow\downarrow}}-1}.
\end{equation}
Explicitly, they are given as
\begin{equation}
\lambda_{k\uparrow}=\lambda_{k\downarrow}=\ln\left(\frac{\sinh^2r_c+\sinh^2r_s+2}{\sinh^2r_c+\sinh^2r_s}\right).
\end{equation}
We note that this GGE has no correlation between the spin-up and spin-down subsystems, and hence this is different from Eq.~(\ref{eq:rho_bar}).
In this way, the property of the stationary state depends on the initial entanglement.

\subsection{Spin-charge representation}
In this subsection, we treat the same problem in the spin-charge representation.
As explained in Sec.~\ref{sec:EP}, the initial entanglement between the two noninteracting subsystems and the energy degeneracies provide the mechanism of the EP.
It is noted that the presence of the entanglement depends on how to decompose the system into the two subsystems.
In the TL model which is split into the two parts $\uparrow$ and $\downarrow$, a natural choice is the subsystem with the $\uparrow$ component and that with the $\downarrow$ component.
However, there is another choice of decomposition into noninteracting subsystems, that is, the spin component and the charge component.
In the spin-charge representation, the Hamiltonian after the quench is given by
\begin{align}
\hat{H}_1^{(TL)}=\hat{H}_c^{(TL)}+\hat{H}_s^{(TL)},
\label{eq:Ham_sc}
\end{align}
where
\begin{align}
\hat{H}_{c/s}^{(TL)}=\int_{-L/2}^{L/2}\left[\frac{(\partial_x\hat{\theta}_{c/s})^2}{4}+2g\hat{n}^2_{c/s}\right]dx.
\label{c/s}
\end{align}
It is apparent from Eqs.~(\ref{eq:initial}) and (\ref{eq:Ham_sc}) that there is no entanglement between the spin and charge subsystems, and there is no interaction in the Hamiltonian after the quench.
Therefore, the spin and charge subsystems will independently evolve to their own stationary states described by the GGE and there is no EP.
Along this line, Kitagawa et al. \cite{Kitagawa2010, Kitagawa2011} calculated the time evolution of the full-distribution function of the interference contrast by utilizing the spin-charge representation.
The result obtained in the spin-charge representation is equivalent to that in the up-down representation presented in Sec.~\ref{sec:up_down}.

In the spin-charge representation, the post-quench Hamiltonian (\ref{TL_1updown}) is given by Eq.~(\ref{eq:Ham_sc}), and the Hamiltonians $\hat{H}_{c/s}^{(TL)}$ are diagonalized as
\begin{align}
\hat{H}_{c/s}^{(TL)}=\sum_k \omega_k b_k^{c/s \dagger}b_k^{c/s}
\end{align}
Thus the time evolution of the charge component is independent of that of the spin component.
The charge and spin components are initially decoupled (no entanglement), and hence they are independent of each other for any $t>0$,
\begin{align}
|\Psi(t)\rangle=|\Psi_c(t)\rangle \otimes |\Psi_s(t)\rangle.
\end{align}

Since the spin and charge components independently relax to their own stationary states, the stationary state will be given by the GGE of charge and spin components,
\begin{align}
\rho_{\rm sc}^{\rm GGE}=\frac{\exp[-\sum_k(\lambda_k^c b_k^{c \dagger}b_k^c + \lambda_k^s b_k^{s \dagger}b_k^s)]}{\mathrm{Tr}\,\exp[-\sum_k(\lambda_k^c b_k^{c \dagger}b_k^c + \lambda_k^s b_k^{s \dagger}b_k^s)]},
\label{eq:rho_gge_sc}
\end{align}
where $\lambda_k^{c}$ and $\lambda_k^{s}$ are determined by the initial values of $b_k^{c\dagger}b_k^c$ and $b_k^{s\dagger}b_k^s$, respectively:
\begin{equation}
\left\{
\begin{split}
\langle\Psi(0)|b_k^{c\dagger}b_k^c|\Psi(0)\rangle=\frac{1}{e^{\lambda_k^c}-1}, \\
\langle\Psi(0)|b_k^{s\dagger}b_k^s|\Psi(0)\rangle=\frac{1}{e^{\lambda_k^s}-1}.
\end{split}
\right.
\end{equation}
Explicitly, they are given as
\begin{align}
\lambda_k^{c/s}=-\ln\left(\tanh^2 r_{c/s}\right).
\end{align}
We can define an effective inverse temperature governing a long-length scale of the spin component as $\beta_{\mathrm{eff}}:=\lim_{k\rightarrow 0}\lambda_k^{s}/\omega_k$.
We obtain
\begin{equation}
\beta_{\mathrm{eff}}=\frac{2}{g},
\end{equation}
which is the same as the one obtained by Kitagawa et al.~\cite{Kitagawa2011}.
Since the cross-correlation function is solely determined by the spin component (see the next section), its long-distance behavior in the stationary state agrees with the thermal equilibrium curve at the effective inverse temperature $\beta_{\mathrm{eff}}$.

\begin{figure}[t]
\begin{center}
\includegraphics[width=8cm]{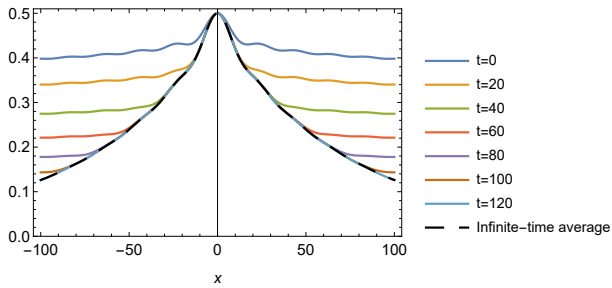}
\caption{Nonequilibrium dynamics of the auto-correlation function for varying times $t$.
The dashed curve represents the infinite-time average calculated by the density matrix~(\ref{eq:rho_bar}).} 
\label{Auto_time}
\end{center}
\end{figure}
\begin{figure}[t]
\begin{center}
\includegraphics[width=8cm]{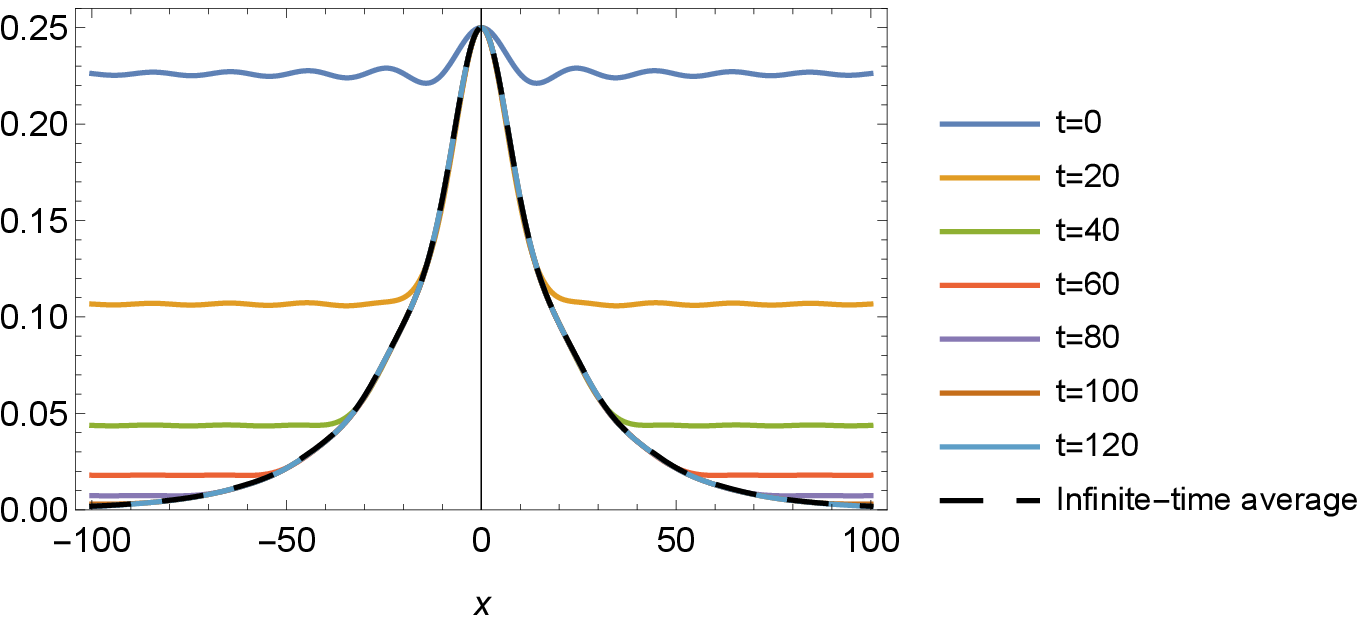}
\caption{Nonequilibrium dynamics of the cross-correlation function for varying times $t$.
The dashed curve represents the infinite-time average calculated by the density matrix~(\ref{eq:rho_bar}).} 
\label{Cross_time}
\end{center}
\end{figure}
\begin{figure}[t]
\begin{center}
\includegraphics[width=7.9cm]{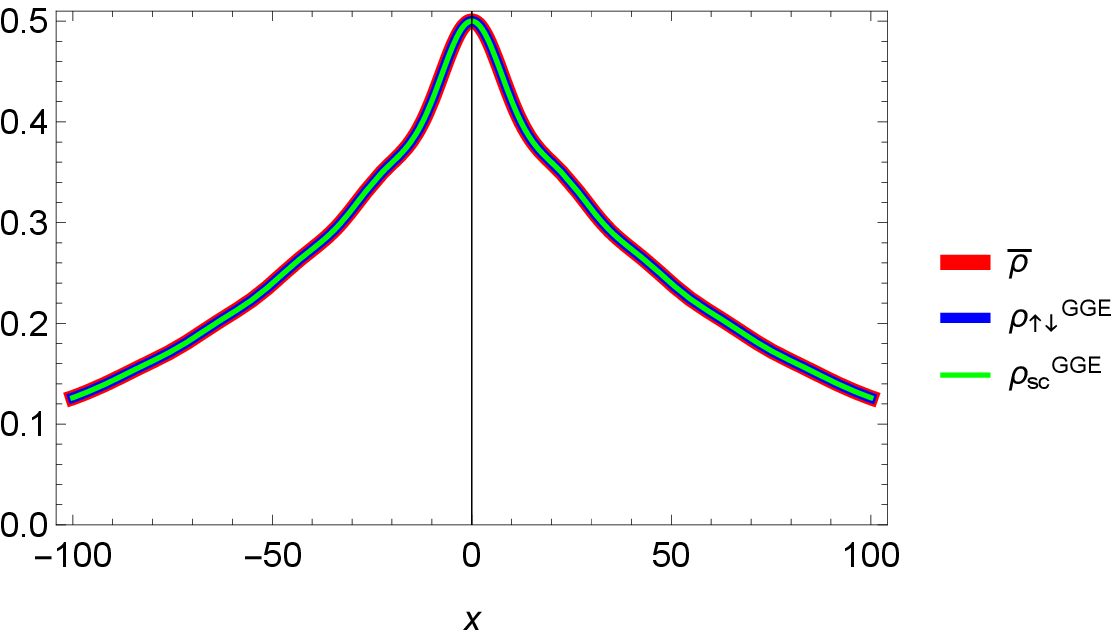}
\caption{Auto-correlation function. The red curve represents the infinite-time average calculated by the density matrix~(\ref{eq:rho_bar}).
The blue curve represents the GGE result in the up-down representation~(\ref{eq:rho_gge_updown}).
The green curve represents the GGE result in the spin-charge representation~(\ref{eq:rho_gge_sc}).}
\label{Auto}
\end{center}
\end{figure}
\begin{figure}[t]
\begin{center}
\includegraphics[width=7.9cm]{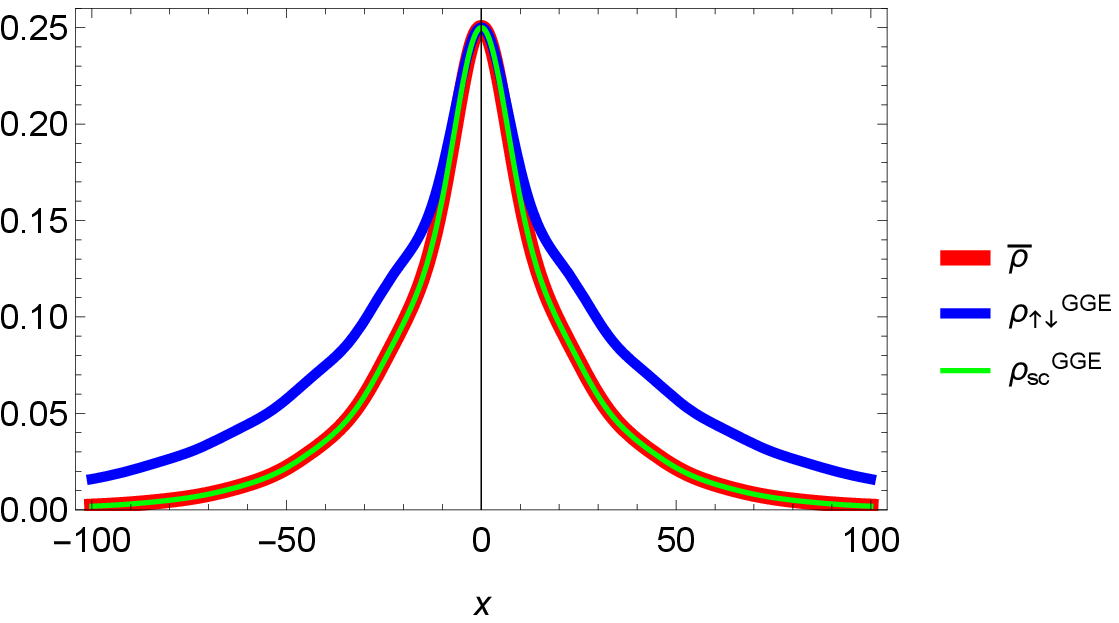}
\caption{{Cross-correlation function. The red curve represents the infinite-time average calculated by the density matrix~(\ref{eq:rho_bar}).
The blue curve represents the GGE result in the up-down representation~(\ref{eq:rho_gge_updown}).
The green curve represents the GGE result in the spin-charge representation~(\ref{eq:rho_gge_sc}).}} 
\label{Cross}
\end{center}
\end{figure}

\subsection{Auto-correlation function and cross-correlation function}
\label{sec:comparison}
We calculate auto-correlation function and the cross-correlation function.
The auto-correlation function is expressed as
\begin{align}
C_{\uparrow}(x)&=\langle\psi^{\dagger}_{\uparrow}(x)\psi_{\uparrow}(0)\rangle
\nonumber \\
&\approx\frac{1}{2}\langle e^{\frac{i}{2}(\theta_c(x)-\theta_c(0))}e^{\frac{i}{2}(\theta_s(x)-\theta_s(0))}\rangle,
\end{align}
where $\langle O\rangle$ denotes the average of $O$ over density matrices such as $\rho(t)=|\Psi(t)\rangle\langle\Psi(t)|$ (exact time evolution), $\bar{\rho}$ (the infinite-time average given in Eq.~(\ref{eq:rho_bar})), $\rho_{\uparrow\downarrow}^{\mathrm{GGE}}$ (GGE in the up-down representation), and $\rho_{\mathrm{sc}}^{\mathrm{GGE}}$ (GGE in the spin-charge representation).
The cross-correlation function is given by
\begin{align}
C_{\uparrow\downarrow}(x)&=\langle\psi^{\dagger}_{\uparrow}(0)\psi^{\dagger}_{\downarrow}(x)\psi_{\uparrow}(x)\psi_{\downarrow}(0)\rangle
\nonumber \\
&\approx\frac{1}{4}\langle e^{{i}(\theta_s(x)-\theta_s(0))}\rangle.
\end{align}
The system size is very large ($N$=$L$=10000), and the interaction strength is set as $g=0.1$.

Figures~\ref{Auto_time} and \ref{Cross_time} show the nonequilibrium time evolution of the auto-correlation and the cross-correlation after a coherent split, respectively.
The black dashed line represents the infinite-time average calculated by using Eq.~(\ref{eq:rho_bar}).
We can see that correlations in the prethermalized state emerge locally and propagate through the system in a light-cone-like evolution, which is consistent with the experiment of Ref.\cite{Langen2013}.

Figures~\ref{Auto} and \ref{Cross} show the comparison of the auto-correlation functions and the cross-correlation functions, respectively, computed by using several different density matrices.
The red curve shows the infinite-time average obtained by using $\bar{\rho}$ in Eq.~(\ref{eq:rho_bar}), the blue curve shows the GGE result in the up-down representation $\rho_{\mathrm{\uparrow \downarrow}}^{\mathrm{GGE}}$, and the green curve shows the GGE result in the spin-charge representation $\rho_{\mathrm{sc}}^{\mathrm{GGE}}$.

As for the auto-correlation function, all the curves agree with each other, indicating that the initial entanglement is not important for the auto-correlation.
As for the cross-correlation function, the GGE curve of $\rho_{\mathrm{sc}}^{\mathrm{GGE}}$ shows an excellent agreement with the infinite-time average, but the GGE curve of  $\rho_{\uparrow\downarrow}^{\mathrm{GGE}}$ deviates significantly from the others.
This deviation is due to the initial entanglement, and clearly shows the EP in the TL model.

Both $\bar{\rho}$ and $\rho_{\mathrm{sc}}^{\mathrm{GGE}}$ nicely describe the prethermalized state.
Thus, the prethermalization in the split TL gases is interpreted as the EP in the ``$\uparrow$'' and ``$\downarrow$'' representation, and it is also interpreted as the prethermalization to the GGEs of two independent subsystems in the spin-charge representation.
These two interpretations are equivalent.

\section{Entanglement Prethermalization in the Presence of with interactions between  two subsystems}

In the TL model, even if there exist interactions between the up and down subsystems,
\begin{equation}
\hat{H}_{\mathrm{int}}=J\int_{-L/2}^{L/2}dx \hat{n}_{\uparrow}(x)\hat{n}_{\downarrow}(x),
\end{equation}
it is decomposed into the purely charge part and the purely spin part as
\begin{equation}
\hat{H}_{\mathrm{int}}=J\int_{-L/2}^{L/2}dx(\hat{n}_c(x)^2-\hat{n}_s(x)^2),
\end{equation}
and hence, the total Hamiltonian $\hat{H}_1'\equiv \hat{H}^{(TL)}_{\uparrow}+\hat{H}^{(TL)}_{\downarrow}+\hat{H}_{\mathrm{int}}$ is written in the form
\begin{equation}
\hat{H}_1'=\hat{H}_c'+\hat{H}_s'=\sum_k(\omega_k^cb_k^{c\dagger}b_k^c+\omega_k^sb_k^{s\dagger}b_k^s),
\label{eq:interacting}
\end{equation}
and the charge and spin components are still decoupled (here, the definitions of $b_k^c$ and $b_k^s$ are different from those in the previous section due to the $\lambda$ term).
This is known as the spin-charge separation in the TL model.
Here, $\omega_k^c$ and $\omega_k^s$ are explicitly given by
\begin{equation}
\left\{
\begin{split}
\omega_k^c=|k|\sqrt{2\left(g+\frac{J}{2}\right)}, \\
\omega_k^s=|k|\sqrt{2\left(g-\frac{J}{2}\right)}.
\end{split}
\right.
\end{equation}
The difference from the non-interacting case is that the dispersion relation $\omega_k^c$ of the boson of the charge component is different from the dispersion relation $\omega_k^s$ of the spin component.
As a result, many degeneracies are lifted compared with the non-interacting case ($\lambda=0$).

However, it is shown that the EP between the up and down subsystems remains nonvanishing even if there exist interactions between them.
In other words, the information of the initial entanglement between the up and down subsystems is not lost even after a long-time evolution, and it affects the prethermalized state.
This is confirmed in the following way.
Because the Hamiltonian after the quench is given by (\ref{eq:interacting}) and $\omega_k^c$ and $\omega_k^s$ are different, the non-thermal steady state of this model is still given by the GGE of charge and spin components,
\begin{align}
&\rho_{sc}^{\mathrm{GGE}}\propto e^{-\sum_k(\lambda_k^cb_k^{c\dagger}b_k^c+\lambda_k^sb_k^{s\dagger}b_k^s)}
\nonumber \\
&=e^{-\sum_k\left[\frac{\lambda_k^c+\lambda_k^s}{2}(b_{k\uparrow}^{\dagger}b_{k\uparrow}+b_{k\downarrow}^{\dagger}b_{k\downarrow})
+\frac{\lambda_k^c-\lambda_k^s}{2}(b_{k\uparrow}^{\dagger}b_{k\downarrow}+b_{k\downarrow}^{\dagger}b_{k\uparrow})\right]}.
\end{align}
The parameters $\lambda_k^c$ and $\lambda_k^s$ are determined from the conserved quantities in the initial state as
\begin{equation}
\langle b_k^{\alpha\dagger}b_k^{\alpha}\rangle=\frac{1}{e^{\lambda_k^{\alpha}}-1},
\quad \alpha=c,s.
\end{equation}
If there were no entanglement between the up and down subsystems in the initial state and $\langle b_{k\uparrow}^{\dagger}b_{k\downarrow}\rangle=\langle b_{k\downarrow}^{\dagger}b_{k\uparrow}\rangle=0$, then we have $\langle b_k^{c\dagger}b_k^c\rangle=\langle b_k^{s\dagger}b_k^s\rangle$ and thus $\lambda_k^c=\lambda_k^s\equiv\lambda_k$.
This implies that the GGE is decoupled in the up and down representation,
$\rho_{sc}^{\mathrm{GGE}}\propto e^{-\sum_k\lambda_kb_{k\uparrow}^{\dagger}b_{k\uparrow}}e^{-\sum_k\lambda_kb_{k\downarrow}^{\dagger}b_{k\downarrow}}$
and the correlation between the up and down subsystems is not important in the state at long times after the quench.
On the other hand, if there is an initial entanglement between the up and down subsystems, $\lambda_k^c\neq\lambda_k^s$ (and $\lambda_k^c-\lambda_k^s$ depends on the strength of the initial entanglement), and the GGE cannot be decomposed into the product of the density matrices of the up and down subsystems.
Therefore, in this case, the correlations between the two subsystems are important in the non-thermal steady state.
In this way, the presence or absence of the initial entanglement strongly affects the steady state even if the two subsystems interact with each other.

\section{Conclusion and Discussion}

We have investigated the prethermalization after a coherent splitting of a one-dimensional Bose gas.
The prethermalization is explained by a combination of the initial entanglement between the two subsystems and energy degeneracies due to symmetries.
If there are energy degeneracies, the initial entanglement persists even after a long-time average.
Because of the importance of the initial entanglement, this prethermalization is called the entanglement prethermalization (EP).

What we find in the EP in the TL model is that the initial entanglement is important for the cross-correlation function, but not for the auto-correlation function.
The prethermalized state is described by a mixture of entangled states, in the up-down representation, which clearly shows the EP in the TL model.
We can also analyze the same problem in the spin-charge representation as done in Ref.~\cite{Kitagawa2011}.
In this representation, the prethermalized state is written as a product of the GGEs for charge and spin components since there is no entanglement between the charge component and the spin component.
Thus, in the TL model, the EP between the up and down components is equivalent to the usual prethermalization without entanglement in the charge and spin components (see also Ref.~\cite{Ikeda2017}).

Moreover, we have found that the initial entanglement still affects the long-time behavior of the system even when the up and down subsystems interact with each other.
This is due to the special feature of the TL model, i.e., the spin-charge separation.

In the previous work~\cite{Kaminishi2015}, EP was studied in the LL model.
Although the TL model is regarded as a low-energy effective theory of the LL model, the EP in the TL model cannot be understood as an approximation of the EP found in the LL model.
The timescale of the EP in the LL model discussed in the previous work is so long that the low-energy approximation is not valid.
Since the dynamics in the TL Hamiltonian gives a good approximation of the dynamics in the LL model in a long but finite timescale, it is expected that the EP in the TL model found in this paper also occurs in the LL model in an intermediate timescale before reaching the true stationary state of the LL model.

\begin{acknowledgments}
This work was supported by JSPS KAKENHI Grants No. JP16J03140, No. JP15K17718, No. 16H06718, and No. JP26287088, 
a Grant-in-Aid for Scientific Research on Innovation Areas ``Topological Quantum Phenomena'' (KAKENHI Grant No. 22103005), and
the Photon Frontier Network Program from MEXT of Japan.
\end{acknowledgments}


\begin{thebibliography}{99}

\bibitem{Neumann1929}
J. v. Neumann, Z. Phys. {\bf 57}, 30 (1929).

\bibitem{Srednicki1994}
M. Srednicki, Phys. Rev. E {\bf 50}, 888 (1994).

\bibitem{Tasaki1998}
H. Tasaki, Phys. Rev. Lett. {\bf 80}, 1373 (1998).

\bibitem{Popescu2006}
S. Popescu, A. J. Short, and A. Winter, Nat. Phys. {\bf 2}, 754 (2006).

\bibitem{Rigol2008}
M. Rigol, V. Dunjko, and M. Olshanii, Nature {\bf 452}, 854 (2008).

\bibitem{Linden2009}
N. Linden, S. Popescu, A. J. Short, and A. Winter, Phys. Rev. E {\bf 79}, 061103 (2009).

\bibitem{Goldstein2010}
S. Goldstein, J. L. Lebowitz, C. Mastrodonato, R. Tumulka, and N. Zangh{\`\i}, Proc. R. Soc, A {\bf 466}, 3203 (2010).

\bibitem{Sato2012}
J. Sato, R. Kanamoto, E. Kaminishi, and T. Deguchi, Phys. Rev. Lett. {\bf 108}, 110401 (2012).

\bibitem{Kinoshita2006}
T. Kinoshita, T. Wenger, and D. S. Weiss, Nature {\bf 440}, 900 (2006).

\bibitem{Hofferberth2007}
S. Hofferberth, I. Lesanovsky, B. Fischer, T. Schumm, and J. Schmiedmayer, Nature {\bf 449}, 324 (2007).

\bibitem{Trotzky2012}
S. Trotzky, Y.-A. Chen, A. Flesch, I. P. McCulloch, U. Schollw\"ock, J. Eisert, and I. Bloch, Nat. Phys. {\bf 8}, 325 (2012).

\bibitem{Kaufman2016}
A. M. Kaufman, M. E. Tai, A. Lukin, M. Rispoli, R. Schittko, P. M. Preiss, M. Greiner, Science {\bf 353}, 794 (2016).

\bibitem{Berges2004}
J. Berges, S. Bors{\'a}nyi, and C. Wetterich, Phys. Rev. Lett. {\bf 93}, 142002 (2004).

\bibitem{Kollar2011}
M. Kollar, F. A. Wolf, M. Eckstein, Phys. Rev. B {\bf 84}, 054304 (2011).

\bibitem{Gring2012} 
M. Gring, M. Kuhnert, T. Langen, T. Kitagawa, and J. Schmiedmayer, Science {\bf 337}, 1318 (2012).


\bibitem{Langen2013ST}
T. Langen, M. Gring, M. Kuhnert, B. Rauer, R. Geiger, D. A. Smith, I. E. Mazets, and J. Schmiedmayer, Eur. Phys. J. Special Topics {\bf 217}, 43 (2013).

\bibitem{Langen2013}
T. Langen, R. Geiger, M. Kuhnert, B. Rauer, and J. Schmiedmayer, Nat. Phys. {\bf 9}, 640 (2013).

\bibitem{Kuhnert2013}
M. Kuhnert, R. Geiger, T. Langen, M. Gring, B. Rauer, T. Kitagawa, E. Demler, D. A. Smith, and J. Schmiedmayer, Phys. Rev. Lett. {\bf 110}, 090405 (2013).

\bibitem{Worm2013}
M. van den Worm, B. C. Sawyer, J. J. Bollinger, and M. Kastner, New J. Phys. {\bf 15}, 083007 (2013).

\bibitem{Langen2015}
T. Langen, S. Erne, R. Geiger, B. Rauer, T. Schweigler, M. Kuhnert, W. Rohringer, I. E. Mazets, T. Gasenzer, and J. Schmiedmayer, Science {\bf 348}, 207 (2015).

\bibitem{Kaminishi2015}
E. Kaminishi, T. Mori, T. N. Ikeda, and M. Ueda, Nat. Phys. {\bf 11}, 1050 (2015).

\bibitem{Ikeda2017}
T. N. Ikeda, T. Mori, E. Kaminishi, and M. Ueda, Phys. Rev. E {\bf 95}, 022129 (2017).

\bibitem{Lieb1963}
E. H. Lieb and W. Liniger, Phys. Rev. {\bf 130}, 1605 (1963).

\bibitem{Kitagawa2010}
T. Kitagawa, S. Pielawa, A. Imambekov, J. Schmiedmayer, V. Gritsev, and E. Demler, Phys. Rev. Lett. {\bf 104}, 255302 (2010).

\bibitem{Kitagawa2011}
T. Kitagawa, A. Imambekov, J. Schmiedmayer, and E. Demler, New. J. Phys. {\bf 13}, 073018 (2011).

\bibitem{Giamarchi2004}
T. Giamarchi, {\sl Quantum physics in one dimension}, Vol. 121. Oxford university press, 2004.

\bibitem{Rigol2007}
M. Rigol, V. Dunjko, V. Yurovsky, and M. Olshanii, Phys. Rev. Lett. {\bf 98}, 050405 (2007).

\bibitem{Rigol2006}
M. Rigol, A. Muramatsu, and M. Olshanii, Phys. Rev. A {\bf 74}, 053616 (2006).

\bibitem{Cazalilla2009}
M. A. Cazalilla, Phys. Rev. Lett. {\bf 97}, 156403 (2006).

\bibitem{Cazalilla2012}
M. A. Cazalilla, A. Iucci and M.-C. Chung, Phys. Rev. E {\bf 85}, 011133 (2012).

\bibitem{Cazalilla2008}
M. Kollar and M. Eckstein, Phys. Rev. A {\bf 78}, 013626 (2008).

\bibitem{Calabrese2011}
P. Calabrese, F. H. L. Essler and M. Fagotti, Phys. Rev. Lett. {\bf 106}, 227203 (2011).

\bibitem{Essler2012}
F. H. L. Essler, S. Evangelisti and M. Fagotti, Phys. Rev. Lett. {\bf 109}, 247206 (2012).

\bibitem{Fagotti2013}
M. Fagotti, Phys. Rev. B {\bf 87}, 165106 (2013).

\bibitem{Bucciantini2014}
L. Bucciantini, M. Kormos, and P. Calabrese, J. Phys. A: Math. Theor. {\bf 47}, 175002 (2014).

\bibitem{Pozsgay2013}
B. Pozsgay, J. Stat. Mech, 07003 (2013).

\bibitem{Fagotti2014}
M. Fagotti, M. Collura, F. H. L. Essler and P. Calabrese, Phys. Rev. B. {\bf 89}, 125101 (2014).

\bibitem{Mierzejewski2014}
M. Mierzejewski, P. Prelovsek and T. Prosen, Phys. Rev. Lett. {\bf 113}, 020602 (2014).

\bibitem{Caux2012}
J.-S. Caux and R. M. Konik, Phys. Rev. Lett. {\bf 109}, 175301 (2012).

\bibitem{Mossel2012}
J. Mossel and J.-S. Caux, New. J. Phys. {\bf 14}, 075006 (2012).

\bibitem{Collura2013}
M. Collura, S. Sotiriadis and P. Calabrese, Phys. Rev. Lett. {\bf 110}, 245301 (2013).

\bibitem{Pozsgay2014}
B. Pozsgay, J. Stat. Mech, 10045 (2014).

\bibitem{Sotiriadis2014}
S. Sotiriadis and P. Calabrese, J. Stat. Mech, 07024 (2014).

\bibitem{Goldstein2015}
G. Goldstein and N. Andrei, Phys. Rev. B {\bf 92}, 155103 (2015).

\end{thebibliography}
\end{document}